\documentclass[aps,pra,twocolumn,superscriptaddress,preprintnumbers]{revtex4-2} % PRL standard class

\usepackage{amsmath,amssymb,amsfonts} 
\usepackage{graphicx} 
\usepackage{hyperref}
\usepackage{bm} 
\usepackage{times} 
\usepackage{xcolor}
\usepackage{physics}
\usepackage{multirow}
\usepackage{pifont}
\usepackage{braket}

\usepackage{tikz}

\definecolor{darkgreen}{HTML}{006400}

\begin{document}

\title{Schmidt-Gauge Non-Local Magic: Representation, Optimality, and Mathematical Properties}
	
\author{Fabio Franchini}
\affiliation{Institut Ruder Bošković, Bijenička cesta 54, Zagreb 10000, Croatia}
	
\author{Salvatore Marco Giampaolo}
\affiliation{Institut Ruder Bošković, Bijenička cesta 54, Zagreb 10000, Croatia}

\begin{abstract}
	While the full non-stabilizerness (magic) of a quantum state contains local, basis-dependent contributions, a non-local formulation based on a minimization over local unitaries isolates the component associated with genuinely non-local correlations. 
	Within such a framework, the Schmidt-gauge formulation of non-local magic provides a direct connection between genuinely non-local non-stabilizer correlations and the entanglement spectrum of quantum many-body states.
	Building on the exact Walsh--Hadamard representation introduced in our accompanying Letter, we develop the mathematical theory associated with this formulation.
	We extend the formalism to arbitrary bipartitions, prove the exactness of the Schmidt gauge for arbitrary $1\times N$ bipartitions, and derive general analytical properties of Schmidt-gauge non-local magic, including entanglement bounds, selection rules, and exact relations with the moments of the normalized Walsh spectrum.
	This representation also provides an interpretation of
	Schmidt-gauge non-local magic as a logarithmic inverse
	participation ratio normalized by a universal harmonic baseline,
	thereby relating it to excess delocalization in Walsh space.
	These results demonstrate that the Walsh--Hadamard representation reveals an underlying discrete harmonic structure that is hidden in the original spectral formulation and provides considerably more than an equivalent expression for Schmidt-gauge non-local magic.
	Rather, it furnishes the natural mathematical framework for its analytical investigation, placing the theory within the broader context of discrete harmonic analysis.
\end{abstract}

\preprint{RBI-ThPhys-2026-29}
	
\maketitle

%%%%%%%%%%%%%%%%%%%%%%%%%%%%%%%%%%%%%%%%%%%%%%%%%%
%%%%%%%%%%%%%%%%%%%%%%%%%%%%%%%%%%%%%%%%%%%%%%%%%%
%%%%%%%%%%%%%%%%%%%%%%%%%%%%%%%%%%%%%%%%%%%%%%%%%%
%%%%%%%%%%%%%%%%%%%%%%%%%%%%%%%%%%%%%%%%%%%%%%%%%%
%%%%%%%%%%%%%%%%%%%%%%%%%%%%%%%%%%%%%%%%%%%%%%%%%%

\section{Introduction}
\label{Sec:1}

Resource theories provide a powerful framework for characterizing non-classical features of quantum systems that are relevant for quantum information processing and reaching quantum advantage~\cite{Coecke2016, Chitambar2019}.
Among them, non-stabilizerness, also known as ``{\it magic}'', has emerged as a fundamental resource within the stabilizer formalism for states and operators, playing a central role in universal fault-tolerant quantum computation~\cite{BravyiKitaev2005, Campbell2017, Howard2014, Veitch2014}.
More recently, it has also proven to be a remarkably sensitive probe of quantum many-body systems~\cite{Oliviero2022, Odavic2023, Korbany2025, Liu2026}, providing valuable insights into quantum criticality~\cite{Tarabunga2023}, scrambling~\cite{Turkeshi2025}, topological phases~\cite{Catalano2026}, and a variety of many-body phenomena that are only partially captured by conventional entanglement-based diagnostics as in Ref.~\cite{Catalano2025}.
While standard magic measures successfully quantify the total non-stabilizerness of a quantum state~\cite{Heinrich2019, Leone2022, Haug2023}, they generally combine contributions originating from local degrees of freedom with genuinely non-local quantum correlations.
Disentangling these two contributions is also essential for identifying basis independent non-stabilizerness and for clarifying its relation to entanglement, many-body correlations, and computationally relevant resources.
This has motivated the introduction of non-local magic, whose purpose is precisely to isolate the genuinely non-local component of non-stabilizerness.
Several complementary formulations of this concept have recently been proposed~\cite{Cao2025, Korbany2025, Timsina2025}, reflecting the growing interest of the community to understand and classify genuinely non-local quantum resources beyond conventional entanglement measures.

Within quantum resource theory, a well-defined notion of non-locality emerges for bipartite pure states, as exemplified for instance by the entanglement entropy case.
Within such a framework, non-local magic has been defined as the amount of non-stabilizerness that cannot be removed by unitary operations restricted to each partition~\cite{Cao2025}.
Although minimizing magic in this way quickly becomes a numerically intractable problem even for moderate-size systems, it was observed in~\cite{Cao2025,Andreadakis2026} that the Schmidt-gauge representative appears to achieve the sought minimum.
In particular, Cao \emph{et al.}~\cite{Cao2025} derived an exact spectral representation of the Schmidt-gauge non-local magic, showing that it depends exclusively on the entanglement spectrum. 
This result is particularly significant because it reveals that the Schmidt-gauge functional admits a purely spectral characterization, providing a compact and elegant description of non-local magic. 
At the same time, however, the resulting expression conceals a rich mathematical structure that is not immediately apparent from its original form, making general analytical results difficult to obtain directly.
More generally, the versatility of the Schmidt-gauge formulation has subsequently been demonstrated in a variety of physical settings, ranging from quantum many-body systems~\cite{Liu2026,Qian2025} to relativistic quantum fields
\cite{Cepollaro2025}, holography and quantum gravity~\cite{Cao2024}, while alternative formulations of non-local magic have also begun to emerge~\cite{Timsina2025,Korbany2025}.
Developing a mathematical framework that makes the hidden structure of the spectral representation explicit is therefore essential for uncovering the analytical properties of Schmidt-gauge non-local magic.

In the accompanying Letter~\cite{letter}, we addressed this problem by introducing an equivalent representation of the Schmidt-gauge non-local magic in terms of Walsh--Hadamard autocorrelations of the entanglement spectrum.
Beyond providing an equivalent expression for the previously known spectral formula, this representation exposes the discrete harmonic structure underlying Schmidt-gauge non-local magic, revealing mathematical features that remain hidden in the original formulation.
This naturally raises a broader mathematical question: to what extent can this harmonic structure be exploited to derive exact analytical results and uncover general properties of Schmidt-gauge non-local magic?
The purpose of the present work is to answer this question by developing the mathematical framework underlying the Walsh--Hadamard representation and exploring its analytical consequences in a systematic way.
Specifically, we first extend the Walsh--Hadamard formalism to arbitrary bipartitions, while the equivalence with the previously known spectral formulation is established in the Supplemental Material of Ref.~\cite{letter}.
We then establish the exactness of the Schmidt gauge for arbitrary $1\times N$ bipartitions, thereby providing a rigorous analytical result that complements the more general numerical evidence reported in Ref.~\cite{letter}.
Finally, we exploit the harmonic structure of the Walsh representation to derive several exact analytical properties of Schmidt-gauge non-local magic, including entanglement bounds, selection rules, and an exact relation with the moments of the normalized Walsh spectrum.
Taken together, these results show that the Walsh--Hadamard representation provides considerably more than an alternative expression for Schmidt-gauge non-local magic.
Rather, it provides a natural mathematical framework in which the structural properties of the measure emerge transparently and can be investigated using the tools of discrete harmonic analysis~\cite{Beauchamp1975,Terras1999}.

The paper is organized as follows.
Section~\ref{Sec:2} introduces the Walsh--Hadamard representation.
Subsection~\ref{SubSec:2.1} derives the formalism for symmetric bipartitions and relates it to the previously known spectral representation, while Subsection~\ref{SubSec:2.2} extends the construction to arbitrary bipartitions.
Section~\ref{Sec:3} establishes the exactness of the Schmidt gauge for arbitrary $1\times N$ bipartitions.
Subsection~\ref{SubSec:3.1} considers local unitary transformations acting exclusively on the larger subsystem, whereas Subsection~\ref{SubSec:3.2} extends the proof to arbitrary local unitary transformations.
Section~\ref{Sec:4} develops the analytical consequences of the Walsh framework.
Specifically, Subsection~\ref{SubSec:4.1} establishes entanglement bounds, Subsection~\ref{SubSec:4.2} derives selection rules, and Subsection~\ref{SubSec:4.3} discusses the connection with the moments of the normalized Walsh spectrum.
Finally, Section~\ref{Sec:5} summarizes the main results and discusses possible future developments.

 %%%%%%%%%%%%%%%%%%%%%%%%%%%%%%%%%%%%%%%%%%%%%%%%%%
 %%%%%%%%%%%%%%%%%%%%%%%%%%%%%%%%%%%%%%%%%%%%%%%%%%
 %%%%%%%%%%%%%%%%%%%%%%%%%%%%%%%%%%%%%%%%%%%%%%%%%%
 %%%%%%%%%%%%%%%%%%%%%%%%%%%%%%%%%%%%%%%%%%%%%%%%%%
 %%%%%%%%%%%%%%%%%%%%%%%%%%%%%%%%%%%%%%%%%%%%%%%%%%
 
 \section{Walsh--Hadamard Representation of Schmidt-Gauge Non-local Magic}
\label{Sec:2}

%%%%%%%%%%%%%%%%%%%%%%%%%%%%%%%%%%%%%%%%%%%%%%%%%%
%%%%%%%%%%%%%%%%%%%%%%%%%%%%%%%%%%%%%%%%%%%%%%%%%%
%%%%%%%%%%%%%%%%%%%%%%%%%%%%%%%%%%%%%%%%%%%%%%%%%%

\subsection{Symmetric bipartitions}
\label{SubSec:2.1}

We begin by considering a bipartite pure state whose two subsystems contain the same number $m$ of qubits. 
Its Schmidt decomposition can be written as
\begin{equation}
	|\psi\rangle = \sum_{\alpha=1}^{\chi} \sqrt{\lambda_\alpha}\, |u_\alpha\rangle_A |v_\alpha\rangle_B, \label{eq:schmidt_decomposition}
\end{equation}
where
\begin{equation}
	\lambda_\alpha\ge0, \qquad \sum_{\alpha=1}^{\chi}\lambda_\alpha=1, \qquad	\chi\le2^m.
\end{equation}
Since the Schmidt vectors form orthonormal bases for the supports of the reduced density matrices, they can always be completed to orthonormal bases of the full Hilbert spaces. 
Consequently, local unitary transformations exist that map them onto the computational basis.
After ordering the Schmidt eigenvalues in descending order and extending the spectrum with zeros whenever $\chi<2^m$, one obtains the Schmidt-gauge representative
\begin{equation}
	|\psi_{\rm Sch}\rangle = \sum_{x\in\mathbb F_2^m}
	\sqrt{\lambda_x}\, |x\rangle_A|x\rangle_B,
	\label{eq:schmidt_gauge}
\end{equation}
where $x=(x_1,\ldots,x_m)$ denotes a binary string and $\mathbb F_2^m$ is the $m$-dimensional vector space over the binary field.
Extending the Schmidt spectrum with zeros whenever $\chi<2^m$ allows all subsequent sums to be written over the full binary space $\mathbb F_2^m$.
The quantity studied throughout this work is the second-order stabilizer R\'enyi entropy evaluated on the Schmidt-gauge representative. 
For an arbitrary pure state $|\phi\rangle$ of $L$ qubits, the second-order stabilizer Rényi entropy is defined by~\cite{Leone2022}
\begin{equation}
	M_2(|\phi\rangle) = -\log_2 \left[ 2^{-L} \sum_{P\in\mathcal P_L} \langle\phi|P|\phi\rangle^4 \right], \label{eq:stabilizer_renyi}
\end{equation}
where $\mathcal P_L$ denotes the set of phase-free Pauli strings.
The Schmidt-gauge non-local magic is therefore defined as 
\begin{equation}
	M_2^{\rm Sch}(\psi) =
	M_2(|\psi_{\rm Sch}\rangle).
	\label{eq:schmidt_magic_definition}
\end{equation}
We now derive an exact expression for Eq.~\eqref{eq:schmidt_magic_definition} that depends only on the Schmidt spectrum. 
As we shall show, this naturally leads to a Walsh--Hadamard representation in terms of a family of binary spectral correlation functions, revealing the harmonic structure underlying the Schmidt-gauge non-local magic.

Using the standard binary representation of Pauli operators~\cite{Gottesman1998, Aaronson2004, NielsenChuang}, every $m$-qubit Pauli operator can be written, up to an overall phase in $\{\pm1,\pm i\}$, as
\begin{equation}
	P(s,k)=X^{s}Z^{k}, \qquad s,k\in\mathbb F_2^m, \label{eq:pauli_binary}
\end{equation}
where
\begin{equation}
	X^{s} = \bigotimes_{j=1}^{m} X_j^{\,s_j}, \qquad Z^{k} = \bigotimes_{j=1}^{m} Z_j^{\,k_j}.
\end{equation}
Since the stabilizer R\'enyi entropy depends only on the fourth power of Pauli expectation values, these global phases play no role and will be omitted throughout the derivation.
The action of these operators on computational-basis states is 
\begin{equation}
	X^{s}|x\rangle = |x\oplus s\rangle,  \qquad	Z^{k}|x\rangle = (-1)^{k\cdot x}|x\rangle,
\end{equation}
where $\oplus$ denotes componentwise addition modulo two and 
\begin{equation}
	k\cdot x = \sum_{j=1}^{m} k_jx_j \pmod 2
\end{equation}
is the binary scalar product. 
Consequently,
\begin{equation}
	\langle x|X^{s}Z^{k}|y\rangle = (-1)^{k\cdot y} \delta_{x,y\oplus s}. 
	\label{eq:matrix_element}
\end{equation}
To evaluate Eq.~\eqref{eq:schmidt_magic_definition}, consider a generic bipartite Pauli operator acting on the two subsystems,
\begin{equation} 
	P=P_A\otimes P_B = X^{s_A}Z^{k_A} \otimes X^{s_B}Z^{k_B}.  
	\label{eq:bipartite_pauli}
\end{equation}
Using the Schmidt-gauge representative in Eq.~\eqref{eq:schmidt_gauge} together with Eq.~\eqref{eq:matrix_element}, the expectation value of a generic bipartite Pauli operator, $P=P_A\otimes P_B$ takes the form
\begin{align}
  \!\!\ \langle P \rangle_{\rm Sch} =  & \!\!\!\! \! \sum_{x,y\in\mathbb F_2^m}  \!\!\!\!  \sqrt{\lambda_x\lambda_y} \langle x|X^{s_A}Z^{k_A}|y\rangle \langle x|X^{s_B}Z^{k_B}|y\rangle \nonumber\\
	& \!\!\!\!\!\!\!\! = \!\!\!\!\! \sum_{x,y\in\mathbb F_2^m}  \!\!\! \sqrt{\lambda_x\lambda_y} (-1)^{(k_A\oplus k_B)\cdot y} \delta_{x,y\oplus s_A} \delta_{x,y\oplus s_B}.
	\label{eq:expectation_pauli} 
\end{align}
The two Kronecker deltas can be simultaneously satisfied only when $s_A=s_B\equiv s$.
Therefore, only Pauli strings inducing the same binary transformation on the two Schmidt registers contribute to the stabilizer sum. 
Introducing the relative binary index $k= k_A \oplus k_B$, the Kronecker deltas further impose $x=y\oplus s$, yielding
\begin{equation}
\! \! \! 	A_s(k)\!  = \! \langle X^{s}Z^{k_A} \! \otimes \!  X^{s}Z^{k_B} \rangle_{\rm Sch}\!  =\! \! \! \sum_{x\in\mathbb F_2^m}\! \!  (-1)^{k\cdot x} \sqrt{\lambda_x\lambda_{x\oplus s}}. \label{eq:As_definition}
\end{equation}
The expectation value is therefore completely characterized by the pair of indices $(s,k)$. 
The remaining task is therefore to determine how many distinct Pauli operators correspond to each pair of indices $(s,k)$.
 
For a fixed pair $(s,k)$, the translation label $s$ is uniquely determined, whereas there are exactly $2^m$ ordered pairs $(k_A,k_B)$ satisfying $k=k_A\oplus k_B$. 
Indeed, one of the two binary vectors can be chosen arbitrarily, while the other is uniquely fixed by the above relation. 
Therefore, every value of $A_s(k)$ appears exactly $2^m$ times in the complete sum over bipartite Pauli operators, yielding
\begin{equation}
	\sum_{P_A,P_B} \langle P_A\otimes P_B\rangle_{\rm Sch}^4 = 2^m \sum_{s,k\in\mathbb F_2^m} A_s(k)^4. \label{eq:Pauli_sum}
\end{equation}
Since the bipartition is symmetric, $L=2m$, substituting Eq.~\eqref{eq:Pauli_sum} into the definition of the stabilizer R\'enyi entropy, Eq.~\eqref{eq:stabilizer_renyi}, we obtain
\begin{equation}
	M_2^{\rm Sch}(\psi) = -\log_2 \left[ 2^{-m} \sum_{s,k\in\mathbb F_2^m}  A_s(k)^4 \right].
	\label{eq:WalshFormula}
\end{equation}
Equation~\eqref{eq:WalshFormula} already shows that the Schmidt-gauge non-local magic depends exclusively on the Schmidt spectrum. 
However, its mathematical structure becomes considerably more transparent after recognizing that the quantities $A_s(k)$ are simply the Walsh--Hadamard transforms of a family of shifted spectral correlation functions.

Let us now introduce the binary spectral correlation functions
\begin{equation}
	f_s(x) = \sqrt{\lambda_x\lambda_{x\oplus s}}.
	\label{eq:shifted_overlap}
\end{equation}
The function $f_s(x)$ encodes the correlations between Schmidt eigenvalues separated by the binary displacement $s$.
Different values of $s$ therefore probe correlations between different regions of the entanglement spectrum.
With this notation, Eq.~\eqref{eq:As_definition} can be rewritten as
\begin{equation}
	A_s(k) = \widehat f_s(k) = \sum_{x\in\mathbb F_2^m} (-1)^{k\cdot x} f_s(x),
	\label{eq:WalshTransform}
\end{equation}
that is, $A_s(k)$ is the Walsh--Hadamard transform of the binary spectral correlation functions $f_s(x)$.
Each binary displacement s therefore defines an independent Walsh spectrum associated with a different shifted overlap of the Schmidt spectrum.
Substituting Eq.~\eqref{eq:WalshTransform} in Eq.~\eqref{eq:WalshFormula} gives 
\begin{equation}
	M_2^{\rm Sch}(\psi)	= -\log_2 \left[ 2^{-m} \sum_{s,k\in\mathbb F_2^m} \widehat f_s(k)^4 \right]. \label{eq:WalshFinalCompact}
\end{equation}
Equation~\eqref{eq:WalshFinalCompact} constitutes the central result of this section.
Its mathematical equivalence to the spectral representation of Ref.~\cite{Cao2025} is established in the Supplemental Material of Ref.~\cite{letter}.
The advantage of eq.~\eqref{eq:WalshFinalCompact} is that it expresses the Schmidt-gauge non-local magic as the normalized fourth moment of the Walsh transforms of a family of binary spectral correlation functions. 
This representation makes the underlying harmonic structure explicit and provides the starting point for all the analytical developments presented in the remainder of this work.

%%%%%%%%%%%%%%%%%%%%%%%%%%%%%%%%%%%%%%%%%%%%%%%%%%
%%%%%%%%%%%%%%%%%%%%%%%%%%%%%%%%%%%%%%%%%%%%%%%%%%
%%%%%%%%%%%%%%%%%%%%%%%%%%%%%%%%%%%%%%%%%%%%%%%%%%

\subsection{Asymmetric bipartitions}
\label{SubSec:2.2}

The derivation presented above assumed, for simplicity, that the two subsystems contain the same number of qubits. 
We now show that the final result remains valid for arbitrary bipartitions.
Without loss of generality, let us assume that subsystem $B$ contains more qubits than subsystem $A$, namely $ m_B>m_A\equiv m$.
However, the Schmidt decomposition still involves at most $2^m$ Schmidt vectors.
Consequently, after fixing the Schmidt basis, the Schmidt-gauge state takes the form
\begin{equation}
	|\psi_{\rm Sch}\rangle = \sum_{x\in\mathbb F_2^m}
	\sqrt{\lambda_x}\, |x\rangle_A|x\rangle_{B_0}|0\rangle_{B_E},
	\label{eq:schmidt_gauge_asymmetric}
\end{equation}
In eq.~\eqref{eq:schmidt_gauge_asymmetric}, the Hilbert space of subsystem $B$ has been decomposed into two subsystems, namely $B_0$ and $B_E$, where $B_0$ contains the $m$ qubits participating in the Schmidt decomposition, while $B_E$ comprises the remaining $m_B-m$ qubits. 
The state of the latter is fixed to the computational basis state $|0\rangle_{B_E}$, reflecting the fact that these degrees of freedom do not contribute to the Schmidt rank and remain completely disentangled from subsystem $A$.

With this decomposition of subsystem $B$, a generic Pauli operator can be written as
\begin{equation}
	P_A\otimes P_B =X^{s_A}Z^{k_A} \otimes X^{s_{B_0}}Z^{k_{B_0}} \otimes X^{s_E}Z^{k_E},
	\label{eq:pauli_asymmetric}
\end{equation}
Using Eq.~\eqref{eq:schmidt_gauge_asymmetric}, we obtain that the expectation value factorizes as
\begin{equation}
	\langle X^{s_A}Z^{k_A} \otimes X^{s_{B_0}}Z^{k_{B_0}} \rangle
	\, \langle 0| X^{s_E}Z^{k_E} |0 \rangle.
	\label{eq:factorization_extra}
\end{equation}
The second factor in eq.~\eqref{eq:factorization_extra} depends exclusively on the ancillary register $B_E$ and is completely independent of the Schmidt spectrum.
Since 
\begin{equation}
	\langle0|X^{s_E}Z^{k_E}|0\rangle = \delta_{s_E,0},
	\label{eq:ancilla_factor}
\end{equation}
only Pauli operators that act diagonally on $B_E$ contribute to the stabilizer sum.
Therefore, every contributing Pauli operator on the effective Schmidt registers $(A,B_0)$ can be combined with an arbitrary diagonal Pauli operator $Z^{k_E}$ acting on $B_E$. 
Since $k_E$ is unconstrained, each contribution is repeated exactly $ 2^{m_B-m} $ times.
The complete Pauli sum therefore becomes
\begin{equation}
	\sum_{P_A,P_B} \langle P_A\otimes P_B\rangle^4 = 2^{m_B-m} \sum_{P_A,P_{B_0}} \langle P_A\otimes P_{B_0} \rangle^4.
	\label{eq:extra_factor}
\end{equation}
Substituting Eq.~\eqref{eq:extra_factor} into the definition of the stabilizer R\'enyi entropy, the additional factor $2^{m_B-m}$ exactly cancels the larger normalization factor $2^{-(m_A+m_B)}$. 
As a consequence, the final expression for the Schmidt-gauge non-local magic is identical to Eq.~\eqref{eq:WalshFormula},
\begin{equation}
	M_2^{\rm Sch}(\psi) = -\log_2 \left[
	2^{-m} \sum_{s,k} A_s(k)^4 \right],
\end{equation}
which is therefore valid for arbitrary bipartitions.

%%%%%%%%%%%%%%%%%%%%%%%%%%%%%%%%%%%%%%%%%%%%%%%%%%
%%%%%%%%%%%%%%%%%%%%%%%%%%%%%%%%%%%%%%%%%%%%%%%%%%
%%%%%%%%%%%%%%%%%%%%%%%%%%%%%%%%%%%%%%%%%%%%%%%%%%
%%%%%%%%%%%%%%%%%%%%%%%%%%%%%%%%%%%%%%%%%%%%%%%%%%
%%%%%%%%%%%%%%%%%%%%%%%%%%%%%%%%%%%%%%%%%%%%%%%%%%

\section{Optimality of the Schmidt gauge for $1\times N$ bipartitions}
\label{Sec:3}

In Sec.~\ref{Sec:2} we introduced the Schmidt-gauge non-local magic as the second-order stabilizer R\'enyi entropy evaluated on the Schmidt representative of a bipartite pure state. 
As shown above, this quantity admits an exact Walsh--Hadamard representation and depends exclusively on the Schmidt (entanglement) spectrum.

A different quantity is the non-local magic, defined as the minimum second-order stabilizer R\'enyi entropy over the entire local-unitary orbit,
\begin{equation}
	M_{2}^{\rm NL}(|\psi\rangle) = \min_{U_A,U_B} M_{2}\!\left[(U_A\otimes U_B)|\psi\rangle\right].
	\label{eq:NLmagic}
\end{equation}
The Schmidt-gauge non-local magic coincides with the non-local magic whenever the Schmidt representative realizes the global minimum of Eq.~\eqref{eq:NLmagic}. 
Whether this property holds for arbitrary bipartitions remains an open problem. 
In this section we prove that the equality holds exactly for every pure state of a $1\times N$ bipartition.

Throughout this section we therefore consider a system of $L=N+1$ qubits partitioned into a single-qubit subsystem $A$ and an $N$-qubit subsystem $B$. 
Any state belonging to the local-unitary orbit of the Schmidt representative can be written as
\begin{equation}
\! \! 	|\psi(U_A,U_B)\rangle \!=\! (U_A\! \otimes\!  U_B)\!  \left( \!  \sqrt{\lambda_0}\,
	|0\rangle_A|0\rangle_B \! +\!  \sqrt{\lambda_1}\, |1\rangle_A|1\rangle_B \! \right)
	\label{eq:local_orbit_1N}
\end{equation}
where $\lambda_0,\lambda_1\ge0$, $\lambda_0+\lambda_1=1$, $U_A\in\mathrm U(2)$, and  \mbox{$U_B\in\mathrm U(2^N)$}.
Our main result is that, for every pure state of a $1\times N$ bipartition, the Schmidt representative minimizes the second-order stabilizer R\'enyi entropy over the entire local-unitary orbit. Consequently,
\begin{equation}
	M_2^{\rm NL}(|\psi\rangle) = M_2^{\rm Sch}(|\psi\rangle).
	\label{eq:optimality_1N}
\end{equation}
The proof relies on the fact that the Schmidt rank of a $1\times N$ bipartition cannot exceed two. 
Consequently, the dependence on the $N$-qubit subsystem can always be expressed in terms of two orthonormal states together with the matrix elements of Pauli operators within the two-dimensional subspace that they span. 
We will first establish a lower-bound for non-local magic and then show that the Schmidt representative saturates it.
Introducing
\begin{equation}
	|\phi_0\rangle	= U_B|0\rangle_B, \qquad |\phi_1\rangle = 	U_B|1\rangle_B,
\end{equation}
Eq.~\eqref{eq:local_orbit_1N} becomes
\begin{equation}
	|\psi(U_A,\phi_0,\phi_1)\rangle \!=\! (U_A\otimes\mathbb I) \!\left(\! \sqrt{\lambda_0}\, |0\rangle_A|\phi_0\rangle \! + \! \sqrt{\lambda_1} |1\rangle_A|\phi_1\rangle \right)
	\label{eq:phi_representation}
\end{equation}
where varying $U_B$ is equivalent to varying over all orthonormal pairs $\{|\phi_0\rangle,|\phi_1\rangle\}$.
Notice that the Schmidt spectrum $\{\lambda_0,\lambda_1\}$ remains invariant under local unitary transformations, so that the optimization acts exclusively on the Schmidt vectors.

The proof naturally separates into two steps. We first solve the optimization problem under the restriction $U_A=\mathbb I$, which contains the essential technical ingredients of the derivation. We then extend the result to an arbitrary local unitary acting on the single-qubit subsystem, thereby completing the proof.

%%%%%%%%%%%%%%%%%%%%%%%%%%%%%%%%%%%%%%%%%%%%%%%%%%
%%%%%%%%%%%%%%%%%%%%%%%%%%%%%%%%%%%%%%%%%%%%%%%%%%
%%%%%%%%%%%%%%%%%%%%%%%%%%%%%%%%%%%%%%%%%%%%%%%%%%

\subsection{Optimization for \texorpdfstring{$U_A=\mathbb{I}$}{UA=I}}
\label{SubSec:3.1}

Under the restriction $U_A=\mathbb I$, the state reduces to
\begin{equation}
	|\psi(\phi_0,\phi_1)\rangle = \sqrt{\lambda_0}\,|0\rangle_A|\phi_0\rangle_B
	+ \sqrt{\lambda_1}\,|1\rangle_A|\phi_1\rangle_B,
	\label{eq:state_phi}
\end{equation}
with $\langle\phi_i|\phi_j\rangle=\delta_{ij}$.
Minimizing the second-order stabilizer R\'enyi entropy is equivalent to maximizing the functional
\begin{equation}
	\mathcal{Q}(\phi_0,\phi_1) = \sum_{P\in\mathcal P_L} \langle\psi(\phi_0,\phi_1)|P|\psi(\phi_0,\phi_1)\rangle^4,
	\label{eq:Qfunctional}
\end{equation}
where the sum extends over all phase-free Pauli operators.
It is therefore sufficient to determine how $\mathcal Q(\phi_0,\phi_1)$ depends on the orthonormal pair $\{|\phi_0\rangle,|\phi_1\rangle\}$.

For every Pauli operator $P_B$ acting on subsystem $B$, we introduce the matrix elements
\begin{equation}
\!\!	a_P\!=\!\langle\phi_0|P_B|\phi_0\rangle,
	b_P\!=\!\langle\phi_1|P_B|\phi_1\rangle,
	c_P\!=\!\langle\phi_0|P_B|\phi_1\rangle.
	\label{eq:abc}
\end{equation}
Since the Schmidt subspace is two dimensional, these quantities completely determine the expectation value of every bipartite Pauli operator.
Indeed, for $P=P_A\otimes P_B$, one finds
\begin{align}
	\langle I\otimes P_B\rangle &= \lambda_0 a_P+\lambda_1 b_P, \nonumber\\
	\langle X\otimes P_B\rangle &= 2\sqrt{\lambda_0\lambda_1}\,\mathrm{Re}\,c_P, \nonumber\\
	\langle Y\otimes P_B\rangle &= 2\sqrt{\lambda_0\lambda_1}\,\mathrm{Im}\,c_P, \label{eq:expectation_values}\\
	\langle Z\otimes P_B\rangle &= \lambda_0 a_P-\lambda_1 b_P.
	\nonumber
\end{align}
Substituting Eq.~\eqref{eq:expectation_values} into Eq.~\eqref{eq:Qfunctional} gives
\begin{align}
	\mathcal Q(\phi_0,\phi_1) =\!\!\! \!\! \sum_{P_B\in\mathcal P_N} \!\!\!\!\! \Big[ 
	& (\lambda_0a_P+\lambda_1b_P)^4 + (\lambda_0a_P-\lambda_1b_P)^4 \nonumber\\
	& \!\!\! + 16\lambda_0^2\lambda_1^2 \Big( (\mathrm{Re}\,c_P)^4 + (\mathrm{Im}\,c_P)^4 \Big) \Big].
	\label{eq:Qrestricted}
\end{align}
Using the elementary inequality $ (\mathrm{Re}\, c_P)^4+(\mathrm{Im}\, c_P)^4 \!\le |c_P|^4$,
and expanding the first two terms in Eq.~\eqref{eq:Qrestricted}, we obtain
\begin{align}
	\mathcal Q(\phi_0,\phi_1) \le \sum_{P_B\in\mathcal P_N} \Big[
	& 2\lambda_0^4a_P^4 + 2\lambda_1^4b_P^4 + 12\lambda_0^2\lambda_1^2a_P^2b_P^2 \nonumber\\
	& + 16\lambda_0^2\lambda_1^2|c_P|^4 \Big].
	\label{eq:Qbound1}
\end{align}
Applying the inequality $2a_P^2b_P^2\le a_P^4+b_P^4$, Eq.~\eqref{eq:Qbound1} becomes
\begin{equation}
	\begin{aligned}
		\mathcal Q(\phi_0,\phi_1) \le \!\!\!\!\! 	\sum_{P_B\in\mathcal P_N} \!\!\!\!\!
		\Big[\! 
		& \left( 2\lambda_0^4 + 6\lambda_0^2\lambda_1^2 \right) a_P^4\\
		& \!\!\!\!\! \!\!\!\! + \left( 2\lambda_1^4 + 6\lambda_0^2\lambda_1^2 \right) b_P^4 + 16\lambda_0^2\lambda_1^2|c_P|^4 \Big].
	\end{aligned}
	\label{eq:Q_identity_quartic_bound}
\end{equation}
Since $ |a_P|,|b_P|,|c_P|\le1 $, one has $ a_P^4\le a_P^2$, $b_P^4\le b_P^2$, and $|c_P|^4\le|c_P|^2$.
Therefore,
\begin{equation}
	\begin{aligned}
		\mathcal Q(\phi_0,\phi_1) \le \!\!\!\!\! 	\sum_{P_B\in\mathcal P_N} \!\!\!\!\!
		\Big[\! 
		& \left( 2\lambda_0^4 + 6\lambda_0^2\lambda_1^2 \right) a_P^2\\
		& \!\!\!\!\! \!\!\!\! + \left( 2\lambda_1^4 + 6\lambda_0^2\lambda_1^2 \right) b_P^2 + 16\lambda_0^2\lambda_1^2|c_P|^2 \Big].
	\end{aligned}
	\label{eq:Q_identity_linear_bound}
\end{equation}
Using Parseval's identity,
\begin{equation}
	\sum_{P_B\in\mathcal P_N}a_P^2	= \sum_{P_B\in\mathcal P_N}b_P^2 = 	\sum_{P_B\in\mathcal P_N}|c_P|^2 = 2^N,
\end{equation}
we finally obtain
\begin{equation}
	\mathcal Q(\phi_0,\phi_1) \le 2^N \left( 2\lambda_0^4 + 2\lambda_1^4 + 28\lambda_0^2\lambda_1^2 \right).
	\label{eq:Q_identity_final_bound}
\end{equation}
The upper bound is saturated by the Schmidt representative, for which the Schmidt vectors coincide with two distinct computational basis states, denoted by $|\alpha\rangle$ and $|\beta\rangle$.
Every Pauli operator maps a computational basis state onto another computational basis state, up to an overall phase. 
Consequently, its restriction to the two-dimensional Schmidt subspace is necessarily diagonal, off-diagonal, or identically zero.
In the first case, $ |a_P|=|b_P|=1$, while $ c_P=0$, whereas in the second, $a_P=b_P=0$, and $|c_P|=1$.
In the third case all three matrix elements vanish.
Moreover, whenever $c_P\neq0$, one has $ c_P\in\{\pm1,\pm i\}$,
so that $ (\operatorname{Re}c_P)^4+(\operatorname{Im}c_P)^4 = |c_P|^4$. 
Likewise, $a_P^4=a_P^2$, $b_P^4=b_P^2$, $|c_P|^4=|c_P|^2$, and, since $a_P^2=b_P^2$ for every Pauli operator,
$ 2a_P^2b_P^2 =a_P^4+b_P^4$.
Hence every inequality employed in the derivation is saturated.
Therefore, the Schmidt representative realizes the global maximum of $\mathcal Q$, which is equivalent to the global minimum of the second-order stabilizer R\'enyi entropy under the constraint $U_A=\mathbb I$.

%%%%%%%%%%%%%%%%%%%%%%%%%%%%%%%%%%%%%%%%%%%%%%%%%%
%%%%%%%%%%%%%%%%%%%%%%%%%%%%%%%%%%%%%%%%%%%%%%%%%%
%%%%%%%%%%%%%%%%%%%%%%%%%%%%%%%%%%%%%%%%%%%%%%%%%%

\subsection{Optimization for arbitrary $U_A$}
\label{SubSec:3.2}

We now extend the previous result by allowing arbitrary local unitary transformations on subsystem $A$.
The dependence on the local unitary acting on subsystem $B$ has already been reduced to the coefficients $a_P$, $b_P$, and $c_P$.
It therefore remains only to determine how the quantity $\mathcal{Q}$ changes under an arbitrary single-qubit rotation.

For fixed orthonormal states $\ket{\phi_0}$ and $\ket{\phi_1}$, the state can be written as
\begin{equation}
	\ket{\psi(U_A,\phi_0,\phi_1)}= (U_A\otimes I)\ket{\psi(\phi_0,\phi_1)},
\end{equation}
where $\ket{\psi(\phi_0,\phi_1)}$ is given by Eq.~\eqref{eq:state_phi}.
As in the previous subsection, minimizing the second-order stabilizer R\'enyi entropy is equivalent to maximizing
\begin{equation}
\!\!\!\!\!\!\!	\mathcal{Q}(U_A,\phi_0,\phi_1) \! = \!\!\!\! \sum_{P\in\mathcal{P}_L}\!\!\!\! \left( \bra{\psi(U_A,\phi_0,\phi_1)} \! P \! \ket{\psi(U_A,\phi_0,\phi_1)} \right)^4
	\label{eq:Qunrestricted}
\end{equation}
The identity operator on subsystem $A$ is invariant under $U_A$, so that $ \langle I\otimes P_B\rangle = 	\lambda_0 a_P+\lambda_1 b_P$. 
The remaining three Pauli operators transform according to
\begin{equation}
	U_A^\dagger\sigma_\mu U_A = \sum_{\nu=x,y,z} R_{\mu\nu}\sigma_\nu, \qquad \mu=x,y,z,
\end{equation}
where $R\in SO(3)$.
Consequently, their expectation values are obtained by applying the same orthogonal transformation to the vector
\begin{equation}
	\!\!\!\! \mathbf{v}_P\!=\! \left( 2\sqrt{\lambda_0\lambda_1}\,\mathrm{Re}\,c_P,\, 2\sqrt{\lambda_0\lambda_1}\,\mathrm{Im}\,c_P,\, \lambda_0 a_P\!-\!\lambda_1 b_P \right)
\end{equation}
namely,
\begin{equation}
	\sum_{\mu=x,y,z}
	\langle\sigma_\mu\otimes P_B\rangle^4
	=
	\sum_{\mu=x,y,z}
	\left(
	\sum_{\nu=x,y,z}
	R_{\mu\nu}v_{P,\nu}
	\right)^4.
\end{equation}
The problem is therefore reduced to deriving an upper bound for the quartic contribution in Eq.~\eqref{eq:Qunrestricted} and showing that this bound is saturated by the Schmidt representative.

Using the decomposition derived above, the quantity $\mathcal{Q}(U_A,\phi_0,\phi_1)$ can be expressed as
\begin{equation}
	\begin{aligned}
		\mathcal{Q}(U_A,\phi_0,\phi_1) \!=\!\!\!\!\! \sum_{P_B\in\mathcal P_N} \!\!\!
		\Bigg[
		&
		(\lambda_0 a_P+\lambda_1 b_P)^4	\\
		&
		\!\!\! + \!\!\!\!\sum_{\mu=x,y,z} \!\!\! \left( \sum_{\nu=x,y,z} R_{\mu\nu}v_{P,\nu}
		\right)^4  \Bigg].
	\end{aligned}
	\label{eq:Q_rotation}
\end{equation}
Since the first term is independent of $U_A$, the optimization over local unitaries is entirely determined by the second term.
Therefore, for each Pauli operator $P_B$, it is sufficient to determine the maximum value of the second term in the right-hand side of Eq.~\eqref{eq:Q_rotation} over all orthogonal matrices $R\in SO(3)$.
It is worth noting that, for any real vector $\mathbf{x}$,
\begin{equation}
	\sum_{\mu=x,y,z}x_\mu^4
	\le
	\left(
	\sum_{\mu=x,y,z}x_\mu^2
	\right)^2,
\end{equation}
where equality holds if and only if at most one component of $\mathbf{x}$ is nonvanishing.
Moreover, since $R$ is an orthogonal matrix, it preserves the Euclidean norm of $\mathbf{v}_P$, namely
\begin{equation}
	\sum_{\mu=x,y,z} \left( \sum_{\nu=x,y,z} R_{\mu\nu}v_{P,\nu} \right)^2 = \sum_{\nu=x,y,z} v_{P,\nu}^2.
\end{equation}
Therefore,
\begin{equation}
	\sum_{\mu=x,y,z} \left( \sum_{\nu=x,y,z} R_{\mu\nu}v_{P,\nu} \right)^4 \le \left( \sum_{\nu=x,y,z} v_{P,\nu}^2 \right)^2,
\end{equation}
which immediately yields
\begin{equation}
	\begin{aligned}
		\mathcal Q(U_A,\phi_0,\phi_1) \le \!\!\!\!\! \sum_{P_B\in\mathcal P_N} \!\!\!\! \Big[
		& (\lambda_0 a_P+\lambda_1 b_P)^4 \\
		& \!\!\!\!\! \!\!\!\!\!\!\!\! \!\!\!\!\!\!\!\!\!\!+ \big( (\lambda_0 a_P-\lambda_1 b_P)^2 + 4\lambda_0\lambda_1|c_P|^2 \big)^2	\Big].
	\end{aligned}
	\label{eq:Q_UA_bound}
\end{equation}

To proceed, we recall that the coefficients $a_P$, $b_P$, and $c_P$ are not independent.
Since $P_B$ is unitary and $\ket{\phi_0}$ and $\ket{\phi_1}$ are orthonormal, Bessel's inequality implies
$ a_P^2+|c_P|^2\le1$, and $b_P^2+|c_P|^2\le1$.
Under these constraints, the following inequality holds:
\begin{equation}
	\begin{aligned}
\!\!\!\!\!\!\!		&
		(\lambda_0 a_P+\lambda_1 b_P)^4 \!+\! \left[ (\lambda_0 a_P-\lambda_1 b_P)^2 + 	4\lambda_0\lambda_1|c_P|^2
		\right]^2 \le
		\\
\!\!\!\!\!\!\!		&
		\left( 2\lambda_0^4 + 6\lambda_0^2\lambda_1^2 \right)a_P^2 \!+\! \left( 2\lambda_1^4 + 	6\lambda_0^2 \lambda_1^2
		\right)b_P^2 + 	16\lambda_0^2\lambda_1^2|c_P|^2.
	\end{aligned}
	\label{eq:algebraic_bound}
\end{equation}
A proof is provided in Appendix~\ref{app:algebraicInequality}.
Substituting Eq.~\eqref{eq:algebraic_bound} into Eq.~\eqref{eq:Q_UA_bound},
we obtain
\begin{equation}
	\begin{aligned}
		\mathcal{Q}(U_A,\phi_0,\phi_1) \le \!\!\!\!\! \sum_{P_B\in\mathcal P_N} \!\!\!\!\!
		\Big[\!
		&\left( 2\lambda_0^4 + 6\lambda_0^2\lambda_1^2 \right)a_P^2 \\
		& \!\!\!\!\! \!\!\!\!\! \!\!\!\!\!  \!\!\!\!\! + \left( 2\lambda_1^4 + 6\lambda_0^2\lambda_1^2 \right)b_P^2 + 	16\lambda_0^2\lambda_1^2|c_P|^2 \Big].
	\end{aligned}
	\label{eq:Q_UA_linear_bound}
\end{equation}
Using Parseval's identities in Eq.~\eqref{eq:Q_UA_linear_bound}, we obtain
\begin{equation}
	\mathcal{Q}(U_A,\phi_0,\phi_1) \le 2^N \left( 2\lambda_0^4 + 2\lambda_1^4 + 28\lambda_0^2\lambda_1^2 \right).
	\label{eq:Q_arbitrary_UA_final_bound}
\end{equation}
Since $\lambda_0+\lambda_1=1$, the expression in parentheses can equivalently be written as
\begin{equation}
	2\lambda_0^4 + 2\lambda_1^4 + 28\lambda_0^2\lambda_1^2 =
	1 + (\lambda_0-\lambda_1)^4 + 16\lambda_0^2\lambda_1^2.
	\label{eq:Q_bound_equivalent_form}
\end{equation}
Therefore,
\begin{equation}
	\mathcal{Q}(U_A,\phi_0,\phi_1) \le 2^N \left[ 1	+ (\lambda_0-\lambda_1)^4 +16\lambda_0^2\lambda_1^2	\right].
	\label{eq:Q_unrestricted_upper_bound}
\end{equation}
The right-hand side of Eq.~\eqref{eq:Q_unrestricted_upper_bound} is precisely the value attained by the Schmidt representative.
Indeed, for
\begin{equation}
	\ket{\psi_{\mathrm{Sch}}} = \sqrt{\lambda_0}\ket{0}_A\ket{0}_B + \sqrt{\lambda_1}\ket{1}_A\ket{1}_B,
\end{equation}
with the remaining $N-1$ qubits of subsystem $B$ in the state $\ket{0}^{\otimes(N-1)}$, one finds exactly
\begin{equation}
	\mathcal{Q}_{\mathrm{Sch}} 	= 	2^N \left[ 1 + (\lambda_0-\lambda_1)^4 + 16\lambda_0^2\lambda_1^2 \right].
	\label{eq:Q_Schmidt_value}
\end{equation}
Consequently $\mathcal{Q}(U_A,\phi_0,\phi_1) \le  \mathcal{Q}_{\mathrm{Sch}}$
for every $U_A\in U(2)$ and every pair of orthonormal states $\ket{\phi_0}$ and $\ket{\phi_1}$.
Since maximizing $\mathcal{Q}$ is equivalent to minimizing the second-order stabilizer R\'enyi entropy, the Schmidt representative realizes the global minimum over the complete local-unitary orbit.
Hence, for every pure state with a $1\times N$ bipartition,
\begin{equation}
	\min_{U_A,U_B} M_2\!\left[ (U_A\otimes U_B)\ket{\psi} \right] = M_2\!\left( \ket{\psi_{\mathrm{Sch}}} \right).
	\label{eq:Schmidt_global_minimum_1xN}
\end{equation}
Therefore, for a single-qubit subsystem, the Schmidt-gauge non-local magic coincides with the minimum over the complete local-unitary orbit.
Although the numerical evidence presented in Ref.~\cite{letter} strongly suggests that the Schmidt-gauge representative continues to realize the global minimum of the local-unitary optimization, extending this proof to arbitrary bipartitions appears to require qualitatively different ideas. 
We leave this interesting problem for future investigations.

%%%%%%%%%%%%%%%%%%%%%%%%%%%%%%%%%%%%%%%%%%%%%%%%%%
%%%%%%%%%%%%%%%%%%%%%%%%%%%%%%%%%%%%%%%%%%%%%%%%%%
%%%%%%%%%%%%%%%%%%%%%%%%%%%%%%%%%%%%%%%%%%%%%%%%%%
%%%%%%%%%%%%%%%%%%%%%%%%%%%%%%%%%%%%%%%%%%%%%%%%%%
%%%%%%%%%%%%%%%%%%%%%%%%%%%%%%%%%%%%%%%%%%%%%%%%%%

\section{General properties of the Walsh--Hadamard representation}
\label{Sec:4}

%%%%%%%%%%%%%%%%%%%%%%%%%%%%%%%%%%%%%%%%%%%%%%%%%%
%%%%%%%%%%%%%%%%%%%%%%%%%%%%%%%%%%%%%%%%%%%%%%%%%%
%%%%%%%%%%%%%%%%%%%%%%%%%%%%%%%%%%%%%%%%%%%%%%%%%%

\subsection{Upper bound in terms of the entanglement entropy}
\label{SubSec:4.1}

The Walsh--Hadamard representation provides a direct upper bound relating the Schmidt-gauge non-local magic to the second-order Rényi entanglement entropy.
The second-order Rényi entanglement entropy associated with the reduced density matrix $\rho_A$ is defined as
\begin{equation}
	S_2(\rho_A)=-\log_2\!\left(\sum_x\lambda_x^2\right).
	\label{eq:entanglement_definition}
\end{equation}
With this notation, the Schmidt-gauge non-local magic satisfies the upper bound
\begin{equation}
	M_2^{\mathrm{Sch}}(\psi)\le2S_2(\rho_A).
	\label{eq:entanglement_nonlocal}
\end{equation}
To prove this inequality we start from the Walsh--Hadamard representation in eq.~\eqref{eq:WalshFormula}.
The strategy is to derive a lower bound on the fourth-order Walsh moment appearing in Eq.~\eqref{eq:WalshFormula}. 
This follows directly from Parseval's identity and the Cauchy--Schwarz inequality.
Indeed, for every fixed binary shift $s$, Parseval's identity gives
\begin{equation}
 \sum_k A_s(k)^2 = 2^m \sum_x 	\lambda_x \lambda_{x\oplus s}.
\end{equation}
Moreover, applying the Cauchy--Schwarz inequality to the $2^m$ functions $A_s(k)^2$, one obtains
\begin{equation}
\sum_k A_s(k)^4	\ge \frac{1}{2^m} \left( \sum_k A_s(k)^2 \right)^2.
\end{equation}	
Substituting Parseval's identity into the previous equation yields
\begin{equation}
\sum_k A_s(k)^4 \ge 2^m \left( \sum_x \lambda_x\lambda_{x\oplus s}\right)^2.
\end{equation}
Summing over all binary shifts,
\begin{equation}
 \sum_{s,k} A_s(k)^4 \ge 2^m \sum_s
 \left(	\sum_x \lambda_x\lambda_{x\oplus s} \right)^2.
\end{equation}
Since every contribution in the sum over binary shifts is non-negative, the right-hand side is bounded from below by retaining only the term corresponding to the trivial displacement $s=0$,
\begin{equation}
\sum_{s,k} A_s(k)^4 \ge 2^m \left( \sum_x \lambda_x^2 \right)^2.
\end{equation}
Substituting this lower bound into Eq.~\eqref{eq:WalshFormula} immediately gives
\begin{equation}
M_2^{\mathrm{Sch}} \le -\log_2 \left[ \left( \sum_x\lambda_x^2 \right)^2 \right] = 2S_2(\rho_A).
\end{equation}

The above derivation also clarifies the origin of the bound.
The Rényi-2 entropy depends exclusively on the autocorrelation sector associated with the trivial binary shift $s=0$.
The Schmidt-gauge non-local magic, instead, includes additional non-negative contributions from all binary translations.
Consequently, the Rényi-2 entropy captures the minimum harmonic contribution allowed by the Walsh--Hadamard decomposition, while the remaining autocorrelation sectors encode genuinely non-local correlations within the entanglement spectrum.
This provides a direct mathematical link between bipartite entanglement and Schmidt-gauge non-local magic.
Our result complements previous studies showing that structural properties of the entanglement spectrum, such as its flatness and antiflatness, provide computable witnesses
and bounds for magic and non-local magic~\cite{Tirrito2024, Odavic2025, Ebner2026}.

%%%%%%%%%%%%%%%%%%%%%%%%%%%%%%%%%%%%%%%%%%%%%%%%%%
%%%%%%%%%%%%%%%%%%%%%%%%%%%%%%%%%%%%%%%%%%%%%%%%%%
%%%%%%%%%%%%%%%%%%%%%%%%%%%%%%%%%%%%%%%%%%%%%%%%%%

\subsection{Selection rules and harmonic constraints}
\label{SubSec:4.2}

Beyond its connection with entanglement, the Walsh--Hadamard representation also reveals several structural properties that follow directly from the definition of the shifted overlap functions
\begin{equation}
	f_s(x)=\sqrt{\lambda_x\lambda_{x\oplus s}},
\end{equation}
independently of the specific Schmidt spectrum.
These properties provide further insight into the harmonic structure underlying Schmidt-gauge non-local magic.

The first consequence is a simple selection rule. 
By construction the shifted overlap functions satisfy
\begin{equation}
	f_s(x\oplus s)=f_s(x),
\end{equation}
for every binary string $x$. 
Using the definition of the Walsh--Hadamard coefficients,
\begin{equation}
	A_s(k)=\sum_x(-1)^{k\cdot x}f_s(x),
\end{equation}
we can perform the change of variable $x\rightarrow x\oplus s$ to obtain
\begin{align}
	A_s(k)	&=\sum_x(-1)^{k\cdot(x\oplus s)} f_s(x\oplus s) \nonumber\\
	&=(-1)^{k\cdot s}\sum_x(-1)^{k\cdot x}f_s(x) \nonumber\\
	&=(-1)^{k\cdot s}A_s(k).
\end{align} 
Therefore, $A_s(k)=0$, whenever $k\cdot s=1$.
Consequently, for every non-vanishing binary shift $s$, exactly one half of the Walsh coefficients vanish identically. 
The Walsh spectrum of each shifted overlap function is therefore supported only on the binary hyperplane
$ k\cdot s=0$, rather than on the full binary Fourier space.

A second important property follows directly from Parseval's identity. 
For every binary shift $s$, one has
\begin{equation}
	\sum_k A_s(k)^2	= 2^m \sum_x \lambda_x\lambda_{x\oplus s}. \label{eq:ParsevalShift}
\end{equation}
Summing Eq.~\eqref{eq:ParsevalShift} over all binary shifts yields
\begin{equation}
	\sum_{s,k}A_s(k)^2 = 2^m \sum_{s,x} \lambda_x\lambda_{x\oplus s}.
\end{equation}
The double sum appearing on the right-hand side can be evaluated exactly. 
For every fixed $x$, the map $ s\longmapsto x\oplus s $ is a bijection of the binary configuration space onto itself, implying $\sum_s\lambda_{x\oplus s} = \sum_y\lambda_y = 1$.
Therefore, $\sum_{s,x} \lambda_x\lambda_{x\oplus s} = \sum_x\lambda_x = 1$, and consequently
\begin{equation}
	\sum_{s,k}A_s(k)^2 = 2^m.
	\label{eq:WalshSecondMoment}
\end{equation}
Equation~\eqref{eq:WalshSecondMoment} shows that the total second-order Walsh moment is completely determined by the normalization of the Schmidt coefficients and is therefore independent of the quantum state.
Consequently, all the state dependence of the Schmidt-gauge non-local magic originates from how this fixed harmonic weight is redistributed among the allowed Walsh modes. 
Since the Schmidt-gauge non-local magic depends on the fourth-order Walsh moment, it quantifies how this fixed spectral weight is concentrated among the allowed Walsh modes rather than its total norm.

%%%%%%%%%%%%%%%%%%%%%%%%%%%%%%%%%%%%%%%%%%%%%%%%%%
%%%%%%%%%%%%%%%%%%%%%%%%%%%%%%%%%%%%%%%%%%%%%%%%%%
%%%%%%%%%%%%%%%%%%%%%%%%%%%%%%%%%%%%%%%%%%%%%%%%%%

\subsection{Moment interpretation of Schmidt-gauge non-local magic}
\label{SubSec:4.3}

The previous results motivate the introduction of the normalized Walsh coefficients 
\begin{equation}
	W_s(k)=2^{-m/2}A_s(k).
\end{equation}
According to Eq.~\eqref{eq:WalshSecondMoment}, these coefficients satisfy $\sum_{s,k}W_s(k)^2=1$.
Introducing the probability distribution
\begin{equation}
	p_{s,k}=W_s(k)^2,
\end{equation}
the Schmidt-gauge non-local magic can be written as
\begin{equation}
	M_2^{\mathrm{Sch}} =-m-\log_2\!\left(\sum_{s,k}W_s(k)^4\right)
	=H_2(p)-m \label{eq:IPRrepresentation}
\end{equation}
where $H_2(p)=-\log_2\sum_{s,k}p_{s,k}^2$ is the second-order R\'enyi entropy of the normalized Walsh distribution. 
Equivalently, defining $\mathrm{IPR}_2(p)=[\sum_{s,k}p_{s,k}^2]^{-1}$, one obtains
\begin{equation}
	M_2^{\mathrm{Sch}} =\log_2\!\left[\frac{\mathrm{IPR}_2(p)}{2^m}\right].
\end{equation}
Thus, Schmidt-gauge non-local magic measures the harmonic delocalization  of the normalized Walsh spectrum relative to the universal baseline $2^m$.
This participation-ratio interpretation is conceptually related to multifractal-flatness approaches to non-stabilizerness~\cite{Turkeshi2023}. 
In the present construction, however, the relevant probability distribution is defined in Walsh space and is generated by the shifted autocorrelations of the Schmidt spectrum.

If $\operatorname{IPR}_2(p)$ remains close to this baseline, the Schmidt-gauge non-local magic is small. 
For maximally entangled  states, $A_s(k)=\delta_{k,0}$ for every shift $s$, so that
$p_{s,k}=2^{-m}\delta_{k,0}$. 
Consequently, $H_2(p)=m$ and $M_2^{\mathrm{Sch}}=0$.
As discussed perturbatively in Ref.~\cite{letter}, Haar-random states are expected to exhibit only a finite excess harmonic entropy $H_2(p)-m$, resulting in a finite but non-extensive Schmidt-gauge non-local magic. 
Conversely, extensive Schmidt-gauge non-local magic requires $\mathrm{IPR}_2(p)$ to become exponentially larger than $2^m$, corresponding to a harmonic distribution spread over an
exponentially enhanced number of Walsh modes. 
Whether such behavior can be realized in physically relevant states remains an open problem.

In conclusion, the Walsh--Hadamard representation replaces a highly non-linear functional of the entanglement spectrum with a normalized measure of harmonic delocalization in Walsh space.

%%%%%%%%%%%%%%%%%%%%%%%%%%%%%%%%%%%%%%%%%%%%%%%%%%
%%%%%%%%%%%%%%%%%%%%%%%%%%%%%%%%%%%%%%%%%%%%%%%%%%
%%%%%%%%%%%%%%%%%%%%%%%%%%%%%%%%%%%%%%%%%%%%%%%%%%
%%%%%%%%%%%%%%%%%%%%%%%%%%%%%%%%%%%%%%%%%%%%%%%%%%
%%%%%%%%%%%%%%%%%%%%%%%%%%%%%%%%%%%%%%%%%%%%%%%%%%
%%%%%%%%%%%%%%%%%%%%%%%%%%%%%%%%%%%%%%%%%%%%%%%%%%

\section{Conclusions}
\label{Sec:5}

In this work we developed a mathematical framework for the Schmidt-gauge non-local magic based on its Walsh--Hadamard representation.
This formulation recasts the Schmidt-gauge functional in terms of discrete harmonic analysis over the binary hypercube~\cite{Terras1999}, revealing a mathematical structure that remained largely hidden in the original definition.
Rather than simply providing an equivalent representation of Schmidt-gauge non-local magic, the Walsh--Hadamard formalism identifies the fundamental role of binary translations, spectral autocorrelations, and harmonic modes in determining the non-local properties of the entanglement spectrum.

Within this framework, we have shown that the Walsh--Hadamard formalism naturally extends from symmetric to completely arbitrary bipartitions, without requiring any assumptions on the relative dimensions of the two subsystems.
We have also proved that, for every $1\times N$ bipartition, the Schmidt representative achieves the global minimum of the local-unitary optimization defining non-local magic, establishing the exact equivalence between the Schmidt-gauge non-local magic and the original definition introduced in Ref.~\cite{Cao2025}.
Although extending this result to generic bipartitions remains an open problem, the harmonic formulation itself is completely general and provides a natural starting point for further analytical investigations.

The Walsh--Hadamard representation also allows several general mathematical properties of Schmidt-gauge non-local magic to be derived in a unified manner.
In particular, we established a rigorous upper bound in terms of the second-order Rényi entanglement entropy~\cite{Horodecki2009}, showing that extensive bipartite entanglement is a necessary condition for extensive Schmidt-gauge non-local magic.
We further derived exact harmonic selection rules and established an exact relation between Schmidt-gauge non-local magic and the moments of the normalized Walsh spectrum.
Taken together, these results show that the Walsh--Hadamard representation transforms Schmidt-gauge non-local magic from a highly non-trivial optimization problem into an object whose fundamental properties follow directly from the harmonic structure of the entanglement spectrum.

This harmonic perspective also provides a new mathematical interpretation of Schmidt-gauge non-local magic.
Expressed in terms of normalized Walsh amplitudes, Schmidt-gauge non-local magic equals the second-order R\'enyi entropy of the harmonic distribution in excess of the universal baseline $m$, or equivalently the logarithm of its inverse participation ratio normalized by $2^m$.
Within this picture, Schmidt-gauge non-local magic is naturally interpreted as a measure of harmonic delocalization in Walsh space, rather than as a complicated nonlinear functional of the entanglement spectrum.

More broadly, the present work suggests that the study of non-local magic may benefit from methods originating in discrete harmonic analysis, complementing more conventional entanglement-based approaches.
Many of the structural properties derived here depend only on the harmonic organization of the Walsh coefficients and are therefore independent of the microscopic details of the underlying quantum system.
We therefore expect the Walsh--Hadamard formulation to provide a fertile mathematical framework for deriving further analytical results, identifying new symmetry relations, and deepening our understanding of non-local magic and related quantum resources.

\section*{Acknowledgements.}
The authors acknowledge support from the project "Implementation of cutting-edge research and its application as part of the Scientific Center of Excellence for Quantum and Complex Systems, and Representations of Lie Algebras", Grant No. PK.1.1.10.0004, co-financed by the European Union through the European Regional Development Fund - Competitiveness and Cohesion Programme 2021-2027 and from the Croatian Science Foundation (HrZZ) through the project IP-2025-02-1667, Mining the Quantum: Frustration, Disorder, and Devices.

\appendix

\section{Proof of the algebraic inequality}
\label{app:algebraicInequality}

In this Appendix, we prove Eq.~\eqref{eq:algebraic_bound}.
For compactness, let $u=\lambda_0$, $v=\lambda_1=1-u$, $a=a_P$, $b=b_P$, and $t=|c_P|^2$. 
The Bessel constraints preceding Eq.~\eqref{eq:algebraic_bound} read 
\begin{equation}
	a^2+t\leq 1,\qquad b^2+t\leq 1,
	\label{eq:appendixBesselConstraints}
\end{equation}
with $u,v,t\geq0$ and $u+v=1$.

Let $\mathcal{L}$ and $\mathcal{R}$ denote, respectively, the left- and right-hand sides of Eq.~\eqref{eq:algebraic_bound}, and define $\Delta=\mathcal{R}-\mathcal{L}$.
Expanding the left-hand side gives
\begin{align}
	\mathcal{L}
	={}&2u^4 a^4 +2v^4 b^4 +12 u^2 v^2 a^2 b^2
	+8 u^3 v\, a^2 t +8 u v^3 b^2 t
	\nonumber\\
	&-16 u^2 v^2 a b t + 16 u^2 v^2 t^2 .
	\label{eq:expandedLeftHandSide}
\end{align}
For fixed $|a|$, $|b|$, and $t$, the only term depending on the relative sign of $a$ and $b$ is $-16 u^2 v^2 a b t$. 
Therefore, $\mathcal{L}$ is maximal, and hence $\Delta$ is minimal, when $a b \leq 0$. It is consequently sufficient to set
\begin{equation}
	a=x,\quad b=-y,\quad x=|a|\geq0, \quad y=|b|\geq0.
\end{equation}

For fixed $x$ and $y$, $\Delta$ is a concave quadratic polynomial in $t$, whose quadratic coefficient is $-16 u^2 v^2$. 
From Eq.~\eqref{eq:appendixBesselConstraints},
the allowed interval is 
\begin{equation}
	0\leq t\leq t_\star, \qquad t_\star=\min(1-x^2,1-y^2).
\end{equation}
A concave function on a closed interval is bounded from below by the minimum of its endpoint values. 
It is therefore enough to prove
\begin{equation}
	\Delta(0)\geq0, \qquad \Delta(t_\star)\geq0.
\end{equation}

At the first endpoint, direct expansion gives 
\begin{align}
	\Delta(0)
	={}& 2 u^4 x^2 (1-x^2) + 2 v^4 y^2 (1-y^2)
	\nonumber\\
	&+6 u^2 v^2	\left[x^2(1-y^2)+y^2(1-x^2)\right].
	\label{eq:deltaAtZero}
\end{align}
All terms in Eq.~\eqref{eq:deltaAtZero} are non-negative because $0\leq x,y\leq1$. Hence $\Delta(0)\geq0$.

To study the second endpoint, we may assume without loss of generality that $x\geq y$, since the inequality is invariant under the simultaneous exchange
\begin{equation}
	(u,x)\longleftrightarrow(v,y).
\end{equation}
It follows that $t_\star=1-x^2$. 
The case $x=0$ is immediate, so we take $x>0$ and introduce
\begin{equation}
	r=x^2,\qquad z=\frac{y}{x}, \qquad 0<r\leq1,\qquad 0\leq z\leq1.
\end{equation}
A direct rearrangement yields
\begin{equation}
\Delta(t_\star) =2r\left[(1-r)F_0(u,z)+rF_1(u,z)\right],
	\label{eq:deltaBoundaryDecomposition}
\end{equation}
where
\begin{equation}
	F_1(u,z) =v^2 (1-z^2)\left(v^2 z^2 + 3 u^2\right) 	\geq0
	\label{eq:F1Appendix}
\end{equation}
and
\begin{align}
	F_0(u,z) ={}& v^2(2 u -1)(4 u-1)z^2 -8u^2 v^2 z
	\nonumber\\
	&+u^2 (16 u^2-26 u+11).
	\label{eq:F0Appendix}
\end{align}
It remains to prove that $F_0(u,z)\geq0$ for $u,z\in[0,1]$.
At the endpoints of the interval in $z$, one has
\begin{equation}
	F_0(u,0)=u^2\left(16u^2-26u+11\right)\geq0.
\end{equation}
Indeed, the quadratic polynomial $16u^2-26u+11$ has positive
leading coefficient and discriminant
\begin{equation}
	(-26)^2-4\cdot16\cdot11=-28<0,
\end{equation}
and is therefore strictly positive for every real $u$.
At the other endpoint,
\begin{equation}
	F_0(u,1)=(2u-1)^4\geq0.
\end{equation}

When $F_0$ is concave as a function of $z$, its minimum on $[0,1]$ is attained at an endpoint. 
The same conclusion holds when $F_0$ is convex but its stationary point lies outside
$[0,1]$. 
The quadratic coefficient is positive only for $u<1/4$ or $u>1/2$, and the stationary point belongs to $[0,1]$ only for $0\leq u\leq u_c$. 
The only remaining case is
\begin{equation}
0\leq u\leq u_c, \qquad u_c=\frac{3-\sqrt5}{4},
\end{equation}
for which the stationary point
\begin{equation}
z_\star=\frac{4u^2}{(1-2u)(1-4u)}.
\end{equation}
belongs to $[0,1]$. 
Evaluating $F_0$ at this point gives
\begin{equation}
	F_0(u,z_\star) = \frac{u^2N(u)}{(1-2u)(1-4u)},
	\label{eq:F0Stationary}
\end{equation}
where
\begin{equation}
	N(u)
	=112u^4-272u^3+244u^2-92u+11.
\end{equation}
The denominator in Eq.~\eqref{eq:F0Stationary} is non-negative on $[0,u_c]$. 
Moreover,
\begin{equation}
	N''(u)=8(168u^2-204u+61)>0
\end{equation}
throughout this interval, while
\begin{equation}
	N'(u_c)=64-40\sqrt{5}<0.
\end{equation}
Thus $N'(u)<0$ for every $u\in[0,u_c]$, and $N$ is decreasing on this interval. Consequently,
\begin{equation}
	N(u)\geq N(u_c)=14-6\sqrt{5}>0.
\end{equation}
Equation~\eqref{eq:F0Stationary} therefore implies $F_0(u,z_\star)\geq0$. 
Together with the endpoint analysis, this proves
\begin{equation}
	F_0(u,z)\geq0 \qquad \text{for all } u,z\in[0,1].
\end{equation}

Since both $F_0$ and $F_1$ are non-negative, Eq.~\eqref{eq:deltaBoundaryDecomposition} 
gives $\Delta(t_\star)\geq0$. 
We have thus shown that $\Delta$ is non-negative at both endpoints of its allowed interval.
Its concavity in $t$ then implies  
\begin{equation}
	\Delta(t)\geq0
\end{equation}
throughout the domain defined by Eq.~\eqref{eq:appendixBesselConstraints}, completing the proof of Eq.~\eqref{eq:algebraic_bound}.

\end{document}